\documentclass[twocolumn]{aastex631} %,linenumbers

\usepackage{hyperref}
\usepackage{color}

\definecolor{notes}{HTML}{5C9384}
\newcommand{\oiiib}{[\textrm{O}\textsc{iii}]\ensuremath{\lambda5007}}
\newcommand{\oiiia}{[\textrm{O}\textsc{iii}]\ensuremath{\lambda4959}}
\newcommand{\oiiidbl}{[\textrm{O}\textsc{iii}]\ensuremath{\lambda\lambda4959,5007}}
\newcommand{\oiiiauroral}{[\textrm{O}\textsc{iii}]\ensuremath{\lambda4363}}
\newcommand{\oiiiuvfull}{\textrm{O}\textsc{iii}]\ensuremath{\lambda\lambda1660,1666}}

\newcommand{\oiibrightdbl}{[\textrm{O}\textsc{ii}]\ensuremath{\lambda\lambda3727,3729}}
\newcommand{\oiiauroraldbl}{[\textrm{O}\textsc{ii}]\ensuremath{\lambda\lambda7320,7330}}

\newcommand{\oisixthree}{[\textrm{O}\textsc{i}]\ensuremath{\lambda6300}}

\newcommand{\siidbl}{[\textrm{S}\textsc{ii}]\ensuremath{\lambda\lambda6717,6731}}
\newcommand{\siiiauroral}{[\textrm{S}\textsc{iii}]\ensuremath{\lambda6312}}
\newcommand{\siiidbl}{[\textrm{S}\textsc{iii}]\ensuremath{\lambda\lambda9069,9532}}

\newcommand{\niiauroral}{[\textrm{N}\textsc{ii}]\ensuremath{\lambda5755}}
\newcommand{\niidbl}{[\textrm{N}\textsc{ii}]\ensuremath{\lambda\lambda6548,6584}}
\newcommand{\niia}{[\textrm{N}\textsc{ii}]\ensuremath{\lambda6548}}
\newcommand{\niib}{[\textrm{N}\textsc{ii}]\ensuremath{\lambda6584}}

\newcommand{\feii}{[\textrm{Fe}\textsc{ii}]\ensuremath{\lambda4360}}

\newcommand{\ariii}{[\textrm{Ar}\textsc{iii}]\ensuremath{\lambda7138}}

\newcommand{\halpha}{\textrm{H}\ensuremath{\alpha}}
\newcommand{\hbeta}{\textrm{H}\ensuremath{\beta}}
\newcommand{\hgamma}{\textrm{H}\ensuremath{\gamma}}

\newcommand{\banana}{SGAS1723$+$34}
\newcommand{\crab}{SGAS1226$+$21}

\begin{document}

\title{TEMPLATES: Direct Abundance Constraints for Two Lensed Lyman-Break Galaxies}

\correspondingauthor{Brian Welch}
\email{brian.d.welch@nasa.gov}

\author[0000-0003-1815-0114]{Brian Welch}
\affiliation{Department of Astronomy, University of Maryland, College Park, MD 20742, USA} 
\affiliation{Observational Cosmology Lab, NASA Goddard Space Flight Center, Greenbelt, MD 20771, USA} 
\affiliation{Center for Research and Exploration in Space Science and Technology, NASA/GSFC, Greenbelt, MD 20771}

\author[0000-0002-4606-4240]{Grace M. Olivier}
\affiliation{Department of Physics and Astronomy and George P. and Cynthia Woods Mitchell Institute for Fundamental Physics and Astronomy, Texas A\&M University, 4242 TAMU, College Station, TX 77843-4242, US}

\author[0000-0001-6251-4988]{Taylor A. Hutchison}
\affiliation{Observational Cosmology Lab, NASA Goddard Space Flight Center, Greenbelt, MD 20771, USA}

\author[0000-0002-7627-6551]{Jane R.~Rigby}
\affiliation{Astrophysics Science Division, Code 660, NASA Goddard Space Flight Center, 8800 Greenbelt Rd., Greenbelt, MD 20771, USA}

\author[0000-0002-4153-053X]{Danielle A. Berg}
\affiliation{Department of Astronomy, The University of Texas at Austin, 2515 Speedway, Stop C1400, Austin, TX 78712, USA}

\author{Manuel Aravena}
\affiliation{Instituto de Estudios Astrof\'{\i}sicos, Facultad de Ingenier\'{\i}a y Ciencias, Universidad Diego Portales, Avenida Ej\'ercito Libertador 441, Santiago, Chile.}

\author[0000-0003-1074-4807]{Matthew B. Bayliss}
\affiliation{Department of Physics, University of Cincinnati, Cincinnati, OH 45221, USA}

\author[0000-0002-3272-7568]{Jack E. Birkin}
\affiliation{Department of Physics and Astronomy and George P. and Cynthia Woods Mitchell Institute for Fundamental Physics and Astronomy, Texas A\&M University, 4242 TAMU, College Station, TX 77843-4242, US}

\author[0000-0002-8487-3153]{Scott C.\ Chapman}
\affiliation{Department of Physics and Atmospheric Science, Dalhousie University, Halifax, NS, B3H 4R2, Canada}
\affiliation{NRC Herzberg Astronomy and Astrophysics, 5071 West Saanich Rd, Victoria, BC, V9E 2E7, Canada}
\affiliation{Department of Physics and Astronomy,  University of British Columbia, Vancouver, BC, V6T1Z1, Canada}

\author[0000-0003-2200-5606]{H\r{a}kon Dahle} \affiliation{Institute of Theoretical Astrophysics, University of Oslo, P.O. Box 1029, Blindern, NO-0315 Oslo, Norway}

\author[0000-0003-1370-5010]{Michael D.~Gladders}
\affiliation{Kavli Institute for Cosmological Physics, University of Chicago, 5640 South Ellis Avenue, Chicago, IL 60637, USA}

\author[0000-0002-3475-7648]{Gourav Khullar}
\affiliation{Department of Physics and Astronomy, and PITT PACC, University of Pittsburgh, Pittsburgh, PA 15260, USA}

\author[0000-0001-6505-0293]{Keunho J. Kim}
\affiliation{IPAC, California Institute of Technology, 1200 E. California Boulevard, Pasadena, CA 91125, USA}

\author[0000-0003-3266-2001]{Guillaume Mahler}
\affiliation{STAR Institute, Quartier Agora - All\'ee du six Ao\^ut, 19c B-4000 Li\`ege, Belgium}

\author[0000-0001-6919-1237]{Matthew A. Malkan}
\affiliation{Department of Physics and Astronomy, University of California, Los Angeles, 430 Portola Plaza, Los Angeles, CA 90095, USA}

\author[0000-0002-7064-4309]{Desika Narayanan}
\affiliation{Department of Astronomy, University of Florida, 211 Bryant Space Sciences Center, Gainesville, FL 32611 USA}
\affiliation{Cosmic Dawn Center (DAWN), Niels Bohr Institute, University of Copenhagen, Jagtvej 128, K{\o}benhavn N, DK-2200, Denmark}

\author[0000-0001-7946-557X]{Kedar A. Phadke}	
\affiliation{Department of Astronomy, University of Illinois Urbana-Champaign, 1002 West Green St., Urbana, IL 61801, USA}	
\affiliation{Center for AstroPhysical Surveys, National Center for Supercomputing Applications, 1205 West Clark Street, Urbana, IL 61801, USA}

\author[0000-0002-7559-0864]{Keren Sharon}
\affiliation{Department of Astronomy, University of Michigan, 1085 S. University Ave, Ann Arbor, MI 48109, USA}

\author[0000-0002-0786-7307]{J.D.T. Smith}
\affiliation{University of Toledo Department of Physics and Astronomy, Ritter Astrophysical Research Center, Toledo, OH 43606} 

\author[0000-0001-6629-0379]{Manuel Solimano}
\affiliation{Instituto de Estudios Astrof\'{\i}sicos, Facultad de Ingenier\'{\i}a y Ciencias, Universidad Diego Portales, Avenida Ej\'ercito Libertador 441, Santiago, Chile.}

\author[0000-0003-3256-5615]{Justin~S.~Spilker}	
\affiliation{Department of Physics and Astronomy and George P. and Cynthia Woods Mitchell Institute for Fundamental Physics and Astronomy, Texas A\&M University, 4242 TAMU, College Station, TX 77843-4242, US}

\author[0000-0001-7192-3871]{Joaquin D. Vieira}	
\affiliation{Department of Astronomy, University of Illinois Urbana-Champaign, 1002 West Green St., Urbana, IL 61801, USA}	
\affiliation{Department of Physics, University of Illinois Urbana-Champaign, 1110 West Green St., Urbana, IL 61801, USA}	
\affiliation{Center for AstroPhysical Surveys, National Center for Supercomputing Applications, 1205 West Clark Street, Urbana, IL 61801, USA}

\author[0000-0002-0786-7307]{David Vizgan}
\affiliation{Department of Astronomy, University of Illinois Urbana-Champaign, 1002 West Green St., Urbana, IL 61801, USA}

%% Mark off the abstract in the ``abstract'' environment. 
\begin{abstract}

Using integrated spectra for two gravitationally lensed galaxies from the JWST TEMPLATES Early Release Science program, we analyze faint auroral lines, which provide direct measurements of the gas-phase chemical abundance. 
For the brighter galaxy, \banana~ ($z = 1.3293$), we detect the \oiiiauroral, \siiiauroral, and \oiiauroraldbl ~ auroral emission lines, and set an upper limit for the \niiauroral ~line. 
For the second galaxy, \crab ~ ($z = 2.925$), we do not detect any auroral lines, and report upper limits. 
With these measurements and upper limits, we constrain the electron temperatures in different ionization zones within both of these galaxies. 
For \banana, where auroral lines are detected, we calculate direct oxygen and nitrogen abundances, finding an N/O ratio consistent with observations of nearby ($z\sim 0$) galaxies. 
These observations highlight the potent combination of JWST and gravitational lensing to measure faint emission lines in individual distant galaxies and to directly study the chemical abundance patterns in those galaxies.

\end{abstract}

%% Keywords should appear after the \end{abstract} command. 
%% The AAS Journals now uses Unified Astronomy Thesaurus concepts:
%% https://astrothesaurus.org
%% You will be asked to selected these concepts during the submission process
%% but this old "keyword" functionality is maintained in case authors want
%% to include these concepts in their preprints.

%% From the front matter, we move on to the body of the paper.
%% Sections are demarcated by \section and \subsection, respectively.
%% Observe the use of the LaTeX \label
%% command after the \subsection to give a symbolic KEY to the
%% subsection for cross-referencing in a \ref command.
%% You can use LaTeX's \ref and \label commands to keep track of
%% cross-references to sections, equations, tables, and figures.
%% That way, if you change the order of any elements, LaTeX will
%% automatically renumber them.
%%
%% We recommend that authors also use the natbib \citep
%% and \citet commands to identify citations.  The citations are
%% tied to the reference list via symbolic KEYs. The KEY corresponds
%% to the KEY in the \bibitem in the reference list below. 

\section{Introduction} \label{sec:intro}

The chemical abundances of galaxies are set by the nuclear synthesis of elements in stars, and the recycling of that enriched gas through the interstellar medium.
As the enriched gas is recycled, measurements of the relative abundance of various elements can provide insight into the lives and deaths of previous generations of stars. 
While astronomers often make the simplifying assumption that the atomic abundance patterns of galaxies are fixed, these patterns in fact should evolve as galaxies evolve. 

The gold standard method to calculate atomic abundances in warm interstellar gas is the ``direct" method, which utilizes auroral emission lines to measure electron temperature and density in the gaseous nebulae \citep[e.g., ][]{Dinerstein1990}. 
These temperature and density measurements are then used in association with typically much brighter collisionally-excited forbidden line emission to calculate abundances of individual elements relative to hydrogen. 

Application of the direct abundance technique is limited by the fact that auroral emission lines are generally faint, at most a few percent the strength of \hbeta.
In local galaxies and HII regions, auroral line abundance patterns have been well studied despite the intrinsic faintness of these emission lines \citep[e.g.,][]{Berg15,Croxall15,Croxall16,Berg20}.
However at higher redshift, this faintness has made such studies more challenging, and therefore only a handful of auroral line detections were possible at Cosmic Noon ($z \sim 1 - 3$) prior to the launch of JWST \citep{Christensen12,James14,Ly15,Ly2016,Sanders2020,Sanders23_keck}.

The difficulty in measuring auroral lines at high redshift leads many to use empirically-calibrated relationships between strengths of strong emission lines and gas-phase metallicity. While these relations can be useful for studies of faint and distant galaxies, previous studies have noted discrepancies between the strong-line metallicity measurements and direct abundances \citep[e.g., ][]{Stasinska05,Kewley&Ellison08,LopezSanchez12}. Additionally, the strong-line indicators have been primarily calibrated with local galaxies and HII regions. As galaxies evolve, we expect their atomic abundance patterns to change, and thus we expect the empirical calibrations to evolve with redshift. However, the limited number of auroral line detections at Cosmic Noon and beyond have made such in-situ recalibrations difficult.

JWST \citep{Gardner23_jwst} has far better sensitivity in the infrared than previous ground and space based telescopes \citep{Rigby23_jwstperf}, and the NIRSpec instrument delivers exquisite near-infrared spectra \citep{Boker23_nirspec}. 
Indeed, the first JWST science data released in summer 2022 (of the SMACS J0723 Early Release Observations \citep[ERO, ][]{Pontoppidan22_ERO} field) emphatically detected the faint auroral \oiiiauroral~ line in galaxies out to $z \sim 8$  \citep{ArellanoCordova22,Schaerer22,Taylor22,Brinchmann23,Curti23auroral,Katz23,Rhoads23,Trump23,Trussler23}.
Early Release Science data from CEERS have detected the same line out to $z\sim 8.7$ \citep{Sanders23_ceers}.

Three cycle 1 programs were approved to study auroral lines at Cosmic Noon (CECILIA, PID 2593, PI Strom, Co-PI Rudie; AURORA, PID 1914, PI Shapley, Co-PI Sanders; PID 1879, PI Curti).
Multiple auroral lines from nitrogen, sulfur, and oxygen have been detected in a stacked sample at $z\sim 2-3$ from CECILIA, enabling more detailed characterizations of chemical abundances \citep{Strom23_Cecilia_lines}.
Recently, the CECILIA team have also published detections of oxygen and sulfur auroral lines in a single galaxy \citep{Rogers24}. 

In this paper we present JWST NIRSpec spectra of two bright, gravitationally lensed galaxies at Cosmic Noon. The magnifying effect of gravitational lensing has enabled the detection of multiple auroral lines in one of these galaxies, and set upper limits on these lines for the other galaxy.

This paper is organized as follows. Section \ref{sec:data} describes the data and data reduction processes. Section \ref{sec:methods} details how we make our measurements. Section \ref{sec:results} discusses our results, and our conclusions are stated in Section \ref{sec:conclusion}. We assume a flat $\Lambda$CDM cosmology with $H_0 = 70$ and $\Omega_m = 0.3$.

\section{Data} \label{sec:data}

\begin{figure*}[t]
    \centering
    \includegraphics[width=0.45\textwidth]{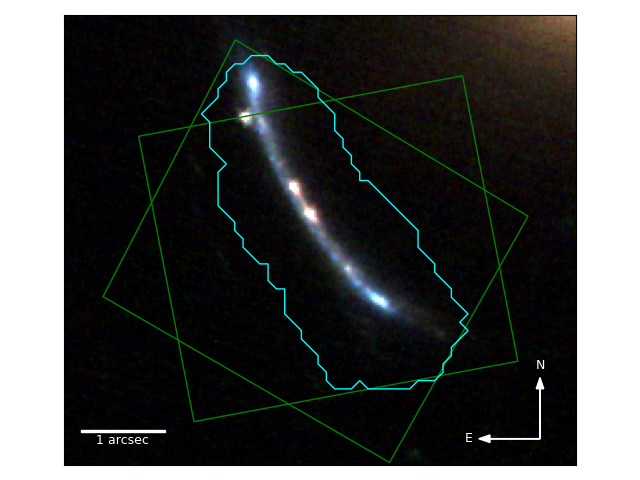}
    \includegraphics[width=0.45\textwidth]{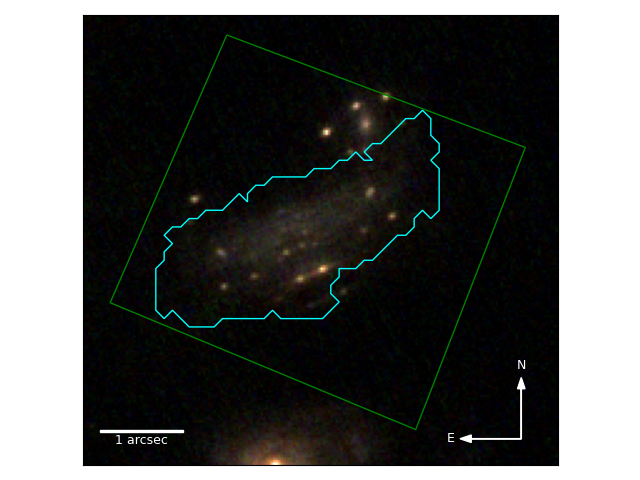}
    \caption{JWST NIRCam color images are shown for \banana~(left) and \crab~(right). The short-wavelength filters F200W, F150W, and F115W are used for the red, green, and blue channels respectively. The green boxes in each image represent the NIRSpec IFU field of view. We display the fields of view of the final cubes after dithering, making each slightly larger than the 3\arcsec by 3\arcsec single-pointing field of view of the IFU. \banana ~ has two sets of observations in the G140H grating (see Section \ref{sec:data}), and we show both in this figure. Cyan contours show the regions used for extraction of the spatially integrated spectra for each target. These regions are based on the SNR of the \oiiib ~line, as described in Section \ref{sec:extract}.}
    \label{fig:fov}
\end{figure*}

The data used in this study were taken as part of the TEMPLATES Director's Discretionary Early Release Science (DD-ERS) observing program with JWST (PID 1355, P.I. Rigby, Co-P.I. Viera). 
TEMPLATES obtained imaging with NIRCam and MIRI, and IFU spectroscopy with NIRSpec and MIRI MRS, targeting four gravitationally lensed arcs \citep{Rigby23_overview}. 
Here we focus on the NIRSpec spectroscopy for two of the four TEMPLATES targets - \banana~ (R.A.17:23:36.406, Dec. $+$34:11:54.69) at $z=1.3293$ \citep{Kubo10,Sharon20,Rigby21} and \crab ~(R.A. 12:26:51.296, Dec. $+$21:52:19.97)at $z=2.925$ \citep{Koester10,Wuyts12,Sharon22templates}. 
These targets are both blue star-forming galaxies with relatively low dust attenuation and sub-solar metallicities, making them the best candidates from TEMPLATES to search for faint auroral line features. 
The two other TEMPLATES targets are dusty star-forming galaxies \citep{Casey14} with near-solar metallicities \citep{Birkin23}.
Auoral lines were not targeted in these two dusty star forming galaxies, and we do not expect to see auroral line emission from such galaxies. Visual inspection of their coadded spectra show no evidence of these lines. 

\banana~ was observed in both the G140H and G395H high spectral resolution (R $\sim$ 3000) gratings.
For this source, G140H covers a rest-frame wavelength range of $\sim 4100$~\AA~ to $\sim 8100$~\AA. The second grating was selected to cover the Paschen-alpha emission line, however we do not use those data in this analysis. 
\crab~ was observed using the G235H grating. 
Further details on the observations can be found in \cite{Rigby23_overview}.

The initial set of observations for \banana~ failed because the telescope drifted off target. 
However, the initial observation in the G140H grating completed successfully prior to this drift. 
Both grating settings were reobserved after this error was discovered.
Thus we have two full observations of \banana~ in the G140H grating, each with the originally allocated exposure time of 4435 seconds. 
We utilize both of these pointings in this analysis in order to maximize signal to noise in our final spectra, resulting in a total usable exposure time of 8870 seconds. 
The IFU field of view from both pointings is shown in Figure \ref{fig:fov}.

The G235H observations of \crab were obtained with an exposure time of 4143 seconds.

The TEMPLATES NIRSpec data reduction is non-trivial. We will provide a brief description here, and refer the interested reader to \cite{Rigby23_overview} for further details. The TEMPLATES team has released a Jupyter Notebook demonstrating how these reductions were performed.\footnote{\url{https://github.com/JWST-Templates/Notebooks}} 
Briefly, we utilize the main JWST reduction pipeline \citep[version 1.11.4, ][]{Bushouse23_pipeine1p11p4}, with several parameters tailored to our specific needs.
We include the \texttt{expand-large-events} flag in the \texttt{Detector1} pipeline to better flag snowballs. 
We run NSClean \citep{Rauscher23_nsclean} to remove correlated read noise from the detector images. 
We utilize the sigma clipping algorithm of \cite{Hutchison23_sigmaclip} to remove any outliers that are not dealt with by the \texttt{Spec3} pipeline's built-in outlier detection step. \cite{Hutchison23_sigmaclip} demonstrated that this combination performs better than either tool alone.

The NIRSpec data only reach down to rest-frame wavelengths of 4100\AA ~for \banana, and 4200\AA ~for \crab. 
The \oiibrightdbl~ lines, which fall outside the NIRSpec coverage, will be useful for determining electron temperatures in the lower ionization region of the galaxy. 
For \banana, this wavelength range was covered by Hubble Space Telescope (HST) WFC3 grism data and Keck ESI data, presented in \cite{Rigby21}.
For the present analysis, for any lines blueward of \hgamma ~in \banana, we use the observed flux measurements reported in \cite{Rigby21}, with our own reddening correction applied as described in Section \ref{sec:redcorr}. 
To account for relative fluxing differences between instruments, we rescale the archival measurements for \banana~ by \hbeta, which is measured by HST WFC3 grism in \cite{Rigby21}. This results in the HST fluxes being multiplied by a factor $1.10 \pm 0.04$. Uncertainties from this rescaling factor are propagated into the resulting line fluxes.
To bolster the confidence of using this rescaling, we check the consistency of the \hbeta/\hgamma ~line ratio between the HST and JWST datasets. We find that the \hbeta/\hgamma ~ratio is consistent within $`\sigma$ between the two datasets ($2.23\pm 0.06$ for JWST, and $2.22\pm 0.2$ for HST).

The Keck ESI spectra were scaled to match the HST grism data based on the \oiibrightdbl ~fluxes. This accounts for potential slit losses from the ESI spectra to first order because the HST grism covers the full width of the source to the sky background limit. 
Second order effects (e.g. from spatial variations perpendicular to the arc) could still affect our measurement. Previous studies have not had sufficient signal to noise to see such variations \citep[e.g.,][]{Florian21}. We assume that these second order effects are small and choose to ignore them for the current analysis.
Additionally, the aperture from which the HST grism and the JWST/NIRSpec IFU spectra are extracted may be slightly different owing to the change in sensitivity between the two instruments. This small aperture difference is corrected by the rescaling process.

For such blue emission lines in \crab, we use fluxes measured from Keck/NIRSPEC reported in \cite{Wuyts12}.
We apply our own measured reddening correction (described in Section \ref{sec:redcorr}) for consistency.
To account for absolute fluxing offsets between the different ground and space based observatories, we normalize the literature reported fluxes based on emission lines covered in both datasets. 
%For \banana, we use the \hbeta ~line for this normalization.
For \crab, \cite{Wuyts12} point out that different levels of atmospheric absorption can affect the absolute fluxing of measured emission lines, and they report the atmospheric transmission calculated for each measured line. 
To account for these differences, we use the \oiiia ~ line as our normalization, as this line has an atmospheric transmission measurement consistent with \oiibrightdbl ~ in the \cite{Wuyts12} data. 
This results in a scaling factor of $0.31 \pm 0.02$. Uncertainties from this rescaling are propagated to the \oiibrightdbl ~line fluxes.
The \oiiia ~and \oiibrightdbl ~lines fall in different filters in this dataset, which could potentially impact the relative fluxing of these lines. 
Additionally, slit losses from the Keck NIRSPEC instrument may impact this calibration, however \cite{Wuyts12} note that the slit covers the entirety of the galaxy, resulting in minimal losses.
We do not attempt to correct for further relative fluxing differences between the archival data and the new JWST observations. 
As a consistency check, we calculate the ratio of the \oiiia/\hbeta ~ lines in both datasets, and find that the ratios are consistent within $1\sigma$ ($7\pm3$ from Keck, and $4.4 \pm 0.2$ for JWST). However we note that these line ratios fall within a single filter in the Keck observations, while the \oiibrightdbl ~lines are observed with a separate filter which may introduce a fluxing offset.

\section{Methods} \label{sec:methods}

\subsection{Extraction of 1D Spectra} \label{sec:extract}

\begin{figure*}[t]
    \centering
    \includegraphics[width=0.95\textwidth]{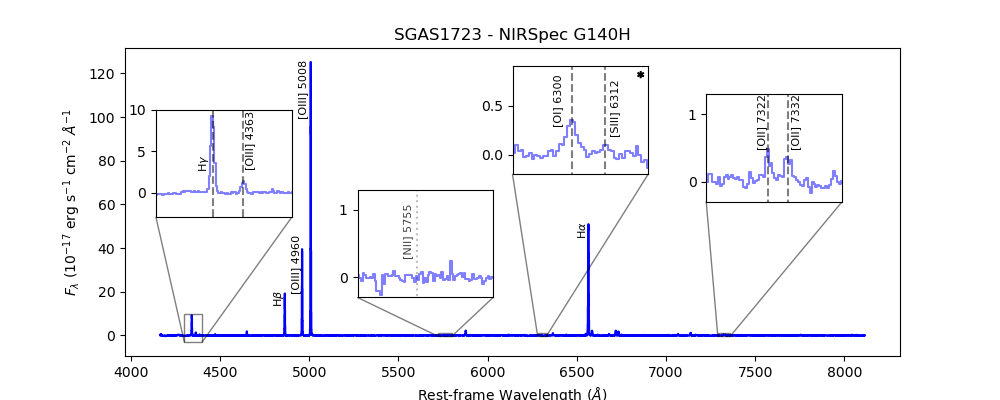} \\
    \includegraphics[width=0.95\textwidth]{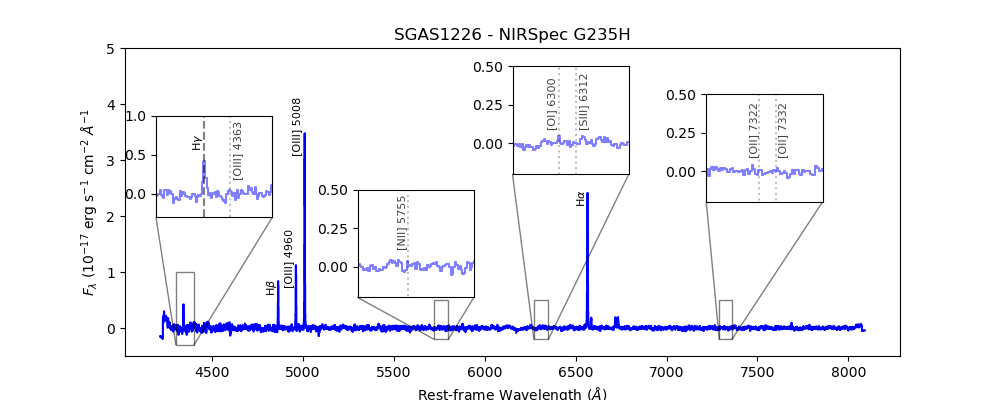}
    \caption{Top: Continuum-subtracted NIRSpec spatially integrated spectrum of \banana, with auroral line regions highlighted in zoomed insets. Three auroral lines are detected, while the fourth (\niiauroral) is not detected. 
    The panel demarcated with an asterisk, containing \oisixthree~and \siiiauroral, shows the spectrum integrated over a smaller aperture, selected to avoid excess noise from pixels at the detector edges (see Appendix \ref{ap:s3local} for details).
    Bottom: Continuum-subtracted NIRSpec spatially integrated spectrum of \crab, with auroral lines highlighted in zoomed insets. No auroral lines are detected in this target.
    In both panels, detected lines are marked with dashed lines, while non-detections are marked with dotted lines.} 
    \label{fig:spec1d}
\end{figure*}

We maximize signal to noise by collapsing the 3D IFU cube down into a single 1D spectrum for each target. 
We create these 1D spectra using masks made as part of the sigma clipping process used in data reduction. 
%The sigma clipping process and the creation of these masks is described in detail in \note{Rigby et al. in prep.} and \note{Hutchison in prep.}. 
Each spaxel included in this mask has signal-to-noise ratio (SNR) $> 3$ for the \oiiib~emission line.
Because the SNR threshold is set based on the brightest emission line, we do not employ any additional aperture corrections when creating our final coadded spectrum. 
Any additional flux scattered outside the aperture area by the instrument PSF will be minimal, particularly in the faint auroral lines on which this analysis is focused. 
We see no evidence of emission from other bright lines extending beyond this extraction region when visually inspecting the 3D IFU cubes.
The final extracted spectra are shown in Figure \ref{fig:spec1d}, with zoomed insets showing the auroral lines covered in each spectrum.

\subsection{Emission Line Measurements} \label{sec:linemeasure}

Line fitting for this paper was done using the JWST TEMPLATES repository of tools, which are publicly available on the TEMPLATES Github repository\footnote{\url{https://github.com/JWST-Templates/jwst_templates}}.
Notebooks used for line fitting and subsequent calculations will be published along with this paper.\footnote{\url{https://github.com/bwelch94/templates-auroral-stacked/}; \dataset[10.5281/zenodo.13697056]{https://zenodo.org/doi/10.5281/zenodo.13697056}} 
Measured line fluxes are reported in Table \ref{tab:lines}.

Significant continuum flux is detected for each of our target galaxies.
To properly measure emission line fluxes, we first measure and subtract the continuum flux for each source. 
This is done by first masking out all detected emission lines, then smoothing the remaining continuum using a boxcar convolution. 
The boxcar size for convolution is chosen to be 100 \AA ~at the source redshift, and we assume a width of 300 km s$^{-1}$ when masking out known emission lines.
This width is chosen prior to fitting, and is intentionally broad to preserve line fluxes. Tests with different mask widths do not significantly change our line flux measurements.
Wavelength ranges that have been masked due to the presence of emission lines are interpolated to provide a smooth estimate of the continuum across the full spectrum. 
The boxcar-smoothed, interpolated spectrum is our estimate of the source continuum level, and is subtracted from the original spectrum prior to fitting emission lines.

We fit the continuum-subtracted emission lines of each spectrum with Gaussian profiles using the \texttt{scipy curve\_fit} function \citep{2020SciPy-NMeth}. 
Line fluxes are calculated by integrating the Gaussian profile. Uncertainties on the Gaussian width and amplitude are propagated to calculate the line flux uncertainties.
Resolved neighboring emission lines (e.g. \oiiia, \oiiib) and partially blended emission lines (e.g. \oiiauroraldbl) are fit simultaneously. 
The \oiiauroraldbl ~ technically consists of four emission lines, with vacuum wavelengths 7320.94\AA, 7322.01\AA, 7331.68\AA, and 7332.75\AA.\footnote{Queried from NIST via \url{https://linelist.pa.uky.edu/atomic/}}
Due to the proximity of the $\sim$7320\AA ~doublet and the $\sim$7330\AA ~doublet, these lines are not resolved in the NIRSpec high resolution gratings. We therefore fit this feature with two Gaussians centered around 7320\AA ~and 7330\AA rest-frame, allowing the wavelength centroids to vary within a flat prior of $\pm 5$\AA ~ in the observed frame. The redshifts of the two lines are not fixed together, however we find that the best-fit redshifts are consistent within $1\sigma$ uncertainties.

We do not fix any doublet line flux ratios while fitting, however we do confirm that our measured flux ratio for \oiiib / \oiiia ~is consistent with the theoretical value given in \cite{storey&zeippen2000}. 
We do not generally fix line widths when fitting our emission lines. However, in the case of the faintest auroral lines, we fix the line widths to what has been measured from related strong lines from the same element, as these will trace the same gas. 
This only notably changes the \oiiauroraldbl ~flux for \banana. As this set of lines has a low SNR, the fitting function favors a wider Gaussian profile than other oxygen lines measured for this target, artificially driving up the flux of the line. We fix the \oiiauroraldbl ~ line widths to the mean velocity width measured from the bright \oiiidbl ~lines. We correct the width for the changing dispersion of the NIRSpec grating, using available dispersion files from the JWST documentation. We note that the \siiiauroral ~line is similarly low SNR, however the best-fit line width matches the predicted line width based on the \siidbl ~line.

Prior to attempting to fit faint lines, we check if there is a significant excess of flux in the line region compared to the surrounding spectral regions.
We first define a region of interest that would contain the emission line. Then we define a separate continuum region in the vicinity of the line region that we do not expect to contain any detectable emission or absorption features. 
We sum the flux densities within the line region.
Then we calculate rolling sums of flux density in the continuum region, with the same width in wavelength as the line region (e.g. if the line region is 10\AA ~wide, we would calculate flux density sums in 10\AA ~bins centered on each resolution element within the continuum region). 
We calculate the median and standard deviation of these rolling flux measurements.
If the line flux sum is less than $3\sigma$ above the median of the rolling fluxes, we consider the line to be undetected.
It can be informative to set upper limits on undetected auroral lines. We set these upper limits at the $3\sigma$ level described above. 

Visual inspection shows the \oiiiauroral~ line may exist just below the detection limit in \crab. We thus attempt a Gaussian fit despite this line being below our standard detection limit. This fit yields a flux of $0.45 \pm 0.53 \times 10^{-17} \textrm{ erg s}^{-1}\textrm{cm}^{-2}\text{\AA}^{-1}$, just below our 3$\sigma$ upper limit of $0.46\times 10^{-17} \textrm{erg s}^{-1}\textrm{cm}^{-2}\text{\AA}^{-1}$, albeit with large uncertainties. The large $1\sigma$ uncertainties from the attempted Gaussian fit are the result of insufficient signal to generate a reliable fit, thus we retain the upper limit calculated using the previously described rolling sum method as our preferred limit.
Additional exposure time could reasonably detect this line in the future.

For the \siiiauroral ~line in \banana, the method described above produces only an upper limit with the original full-galaxy spectrum.
However, looking at the IFU data cube, it appears that a line is present at the expected wavelength, however it is partially cut off by the NIRSpec detector gap.
We therefore perform a separate spectral extraction and local continuum subtraction in an effort to properly measure the flux of this line. 
This process is described in Appendix \ref{ap:s3local}.
We report the flux measured in this smaller aperture with local continuum subtraction in Table \ref{tab:lines}, with an asterisk to denote the atypical measurement.

\begin{table*}[t]
    \centering
    \caption{Emission Line Flux Measurements}
    \begin{tabular}{c c c c c c}
    \hline \hline
        Line & $\lambda_{\textrm{rest}}$ & Flux (S1723) & Intensity $I(\lambda)$ (S1723) & Flux (S1226) & Intensity $I(\lambda)$ (S1226) \\
        \hline 
H$\gamma$ & $4341.68$ &$69.67 \pm 0.97$ & $84.89 \pm 1.19$ &$4.96 \pm 0.53$ & $8.78 \pm 0.93$ \\ 

[O~III]~4363 & $4364.44$ &$11.34 \pm 1.07$ & $13.80 \pm 1.30$ &$< 0.46$ & $< 0.81$ \\ 

H$\beta$ & $4862.68$ &$155.52 \pm 3.42$ & $184.49 \pm 4.06$ &$10.46 \pm 0.50$ & $17.12 \pm 0.82$ \\ 

[O~III]~4960 & $4960.30$ &$325.48 \pm 3.83$ & $384.43 \pm 4.56$ &$14.05 \pm 0.54$ & $22.72 \pm 0.87$ \\ 

[O~III]~5008 & $5008.24$ &$976.50 \pm 5.66$ & $1150.99 \pm 6.85$ &$46.35 \pm 0.65$ & $74.46 \pm 1.04$ \\ 

[N~II]~5756 & $5756.24$ &$< 0.45$ & $< 0.51$ &$< 0.12$ & $< 0.19$ \\ 

[O~I]~6302 & $6302.05$ &$4.09 \pm 0.25$ & $4.80 \pm 0.30$ &$0.00 \pm 0.00$ & $0.00 \pm 0.00$ \\ 

[S~III]~6313 & $6313.80$ &$0.91 \pm 0.22$\tablenotemark{*} & $1.07 \pm 0.25$\tablenotemark{*} &$< 0.11$ & $< 0.16$ \\ 

[N~II]~6550 & $6549.85$ &$5.74 \pm 2.11$ & $6.48 \pm 2.38$ &$0.77 \pm 0.43$ & $1.08 \pm 0.61$ \\ 

H$\alpha$ & $6564.61$ &$503.84 \pm 3.79$ & $568.09 \pm 4.32$ &$39.53 \pm 0.64$ & $55.88 \pm 0.90$ \\ 

[N~II]~6585 & $6585.28$ &$19.39 \pm 2.11$ & $21.85 \pm 2.38$ &$2.33 \pm 0.47$ & $3.29 \pm 0.66$ \\ 

[S~II]~6718 & $6718.29$ &$23.76 \pm 1.08$ & $26.70 \pm 1.22$ &$3.75 \pm 0.32$ & $5.24 \pm 0.45$ \\ 

[S~II]~6733 & $6732.67$ &$17.68 \pm 1.07$ & $19.85 \pm 1.20$ &$3.59 \pm 0.33$ & $5.02 \pm 0.46$ \\ 

[Ar~III]~7138 & $7137.80$ &$12.23 \pm 0.87$ & $13.61 \pm 0.97$  & $< 0.13$ & $< 0.17$ \\

[O~II]~7322 & $7322.01$ &$3.50 \pm 0.45$ & $3.88 \pm 0.50$ &$< 0.09$ & $< 0.12$ \\ 

[O~II]~7332 & $7331.68$ &$3.22 \pm 0.38$ & $3.57 \pm 0.42$ &$< 0.09$ & $< 0.12$ \\ 
        \hline
    \end{tabular}
%    \tablenotetext{a}{Rest-frame vacuum wavelength, in \AA} 
%    \tablenotetext{b}{Observed fluxes reported in units of $10^{-17}$ erg s$^{-1}$ cm$^{-2}$}
%    \tablenotetext{c}{Intensities after reddening correction, in $10^{-17}$ erg s$^{-1}$ cm$^{-2}$}
    \tablenotetext{}{\textbf{Note.} Emission line flux measurements. Columns are: (1) Line identification, (2) Rest-frame vacuum wavelength, in \AA (3) Observed flux for \banana, $10^{-17}$ erg s$^{-1}$ cm$^{-2}$, (4) Dereddened intensity for \banana, $10^{-17}$ erg s$^{-1}$ cm$^{-2}$ (5) Observed flux for \crab, $10^{-17}$ erg s$^{-1}$ cm$^{-2}$ (6) Dereddened intensity for \crab, $10^{-17}$ erg s$^{-1}$ cm$^{-2}$. $3\sigma$ upper limits are given for lines without clear detections. } 
    \tablenotetext{*}{\siiiauroral ~is partially cut off by the detector gap in \banana, thus this line is only measured for half of the galaxy}
    \label{tab:lines}
\end{table*}

\subsection{Temperature, Density, and Abundance Calculations}

\subsubsection{Extinction Correction} \label{sec:redcorr}
We first correct our measured emission line fluxes for dust reddening, both from Milky Way interstellar medium (ISM) and dust within the target galaxy. 
In both cases we use the extinction law of \cite{Cardelli1989}.
The Milky Way ISM reddening correction is made using a 3D dust map from \cite{Bayestar2019dust}, accessed through the \texttt{dustmaps} python package \citep{Green18_dustmaps}. 
At the position of \banana, the Milky Way $E(B-V) = 0.03$, while for \crab ~Milky Way $E(B-V) = 0.00$. 

After accounting for Galactic reddening, we next correct for dust within the target galaxy using the ratios of \halpha ~to \hbeta ~and \hbeta ~to \hgamma. 
We first calculate a predicted \halpha/\hbeta ~and \hbeta/\hgamma ~ratios using $\texttt{PYNEB}$ \citep{pyneb_luridiana15}. 
This initial calculation assumes an electron temperature of $10^4$ K and density of 100 cm$^{-3}$. 
The measured \halpha/\hbeta ~and \hbeta/\hgamma ~ratios are then compared to the predicted ratios to calculate the reddening correction. 
This reddening correction is then applied to our measured emission lines, and a new temperature and density are calculated using the \oiiiauroral/\oiiib ~and \siidbl ~ratios, respectively. 
This process is repeated iteratively until the difference in temperature between iterations is less than 10 K. 
In the case of \crab, no line is detected for \oiiiauroral. For this process, we treat the upper limit as a detection, iteratively measuring the temperature from the flux upper limit until the change is less than 10K. 
This process results in reddening measurements of $E(B-V) = 0.05 \pm 0.01$ for \banana, and $E(B-V) = 0.11 \pm 0.04$ for \crab.

\subsubsection{Electron Temperature and Density Calculations}
For \banana, we have detected the \oiiiauroral, \oiiauroraldbl, and \siiiauroral ~auroral emission lines, and placed an upper limit on the \niiauroral ~auroral line.
Meanwhile \crab ~has yielded only upper limits on each of these auroral lines. 
We calculate electron densities and temperatures from these emission line measurements using the \texttt{PYNEB} function \texttt{getTemDen} \citep{pyneb_luridiana15}.

We first calculate the electron density for each target using the \siidbl ~ line ratio, which is well detected in both galaxies and is sensitive to density.
To account for uncertainties in line flux measurements, we randomly sample a Gaussian distribution with mean and standard deviation given by our line flux and flux uncertainty. We calculate the density using 300 samples, and use the median and standard deviation of the resulting distribution as our density and uncertainty.
Below a density of $\sim100~\textrm{cm}^{-3}$, this ratio does not change considerably as a function of density \citep[see, e.g.][]{Berg18}. In the case of \banana, we are in this regime, leading to an increased uncertainty in our density measurement. For this reason, we choose to use a fiducial density of $100~\textrm{cm}^{-3}$ for our temperature calculations. 
\crab ~is not in this low density regime, so we use our measured value of $\sim 600 ~\textrm{cm}^{-3}$ for future calculations. 

We calculate electron temperature for the highly ionized gas from the \oiiiauroral/\oiiib ~line ratio. For the lower ionization gas we use the \oiiauroraldbl/\oiibrightdbl ~temperature sensitive line ratio.
While we have a detection of \siiiauroral, we do not have the wavelength coverage to measure \siiidbl, which is needed to calculate the electron temperature in the intermediate ionization zone.  
To estimate uncertainties on the temperature measurement, we randomly select line ratios from a Gaussian distribution with mean equal to the measured ratio and standard deviation equal to the uncertainty on the measured ratio. 
We use 300 points randomly sampled from this distribution. 
The physical quantity is then calculated for each sample.
The median of the resulting distribution of physical parameters is taken as our final value, with the standard deviation providing the uncertainty on this value. 

The \oiiauroraldbl ~ temperature calculation relies on a measurement of \oiibrightdbl, which is outside the wavelenth range of the JWST NIRSpec data for both targets. 
We thus use existing measurements from other observatories of this line for both targets. 
As described in Section \ref{sec:data}, \banana~ has \oiibrightdbl ~ fluxes measured by both HST WFC3 grism and Keck ESI, reported in \cite{Rigby21}.

\crab~ has \oiibrightdbl ~blended flux observed by the Keck NIRSPEC instrument, reported in \cite{Wuyts12}. 
As discussed in Section \ref{sec:data}, we rescale the reported \oiibrightdbl~ flux by the \oiiia ~flux, as the reported atmospheric transmission for these two lines is consistent for the ground-based data.

Finally, we apply our own reddening corrections to the rescaled observed fluxes obtained from the literature, as described in Section \ref{sec:redcorr}. 
This results in dereddened intensities of $I(\lambda3727)=109.0\pm 0.8 \times 10^{-17}$ erg s$^{-1}$ cm$^{-2}$, $I(\lambda3729)=153.4\pm 0.8 \times 10^{-17}$ erg s$^{-1}$ cm$^{-2}$ for \banana, and $I(\lambda\lambda3727,3729)=48 \pm 9 \times 10^{-17}$ erg s$^{-1}$ cm$^{-2}$ for \crab.

Our upper limit on the electron temperature for O$^+$ in \crab ~ is extremely low. Because this limit relies on the uncertain rescaling of the ground-based \oiibrightdbl~line fluxes, we conclude that the extremely low O$^+$ temperature is likely erroneous. We include it here for completeness only.

Calculating a temperature from the \siiiauroral ~ line requires a measurement of the \siiidbl, which falls outside the wavelength coverage of the JWST NIRSpec gratings obtained from TEMPLATES. 
We thus estimate the [\textrm{S}\textsc{iii}] temperature using the relation from \cite{G92},
\begin{equation}
    T_e(\textrm{S}^{++}) = 0.83 T_e(\textrm{O}^{++}) + 1700
\end{equation}

For [\textrm{N}\textsc{ii}], we employ our upper limit to calculate a temperature based on the \niiauroral /\niidbl ~ratio limit. We then compare this value to the temperature estimated from $T_e(\textrm{O}^{++})$ using the relation from \cite{G92}, 
\begin{equation} \label{eq:n2o3}
    T_e(\textrm{N}^+) = 0.7 T_e(\textrm{O}^{++}) + 3000
\end{equation}
The \cite{G92} temperature relations equate $T_e(\textrm{O}^+) = T_e(\textrm{N}^+)$. Our measured value for $T_e(\textrm{O}^+)$ is slightly higher than the value calculated based on the measured $T_e(\textrm{O}^{++})$, with a discrepancy of $\sim 2\sigma$. 
Our measured $T_e(\textrm{O}^+)$ is also $\sim 1.2\sigma$ greater than our upper limit on $T_e(\textrm{N}^+)$ from our non-detection of the \niiauroral ~line. We thus choose to use the calculated value of $T_e(\textrm{N}^+)_{\textrm{TT}}$ for further calculations.

\subsubsection{Abundance Calculations}
\label{sec:abundcalc}
We can use our measured electron temperatures to calculate direct elemental abundances within \banana. For \crab ~we only set an upper limit on the oxygen abundance, as no auroral lines are detected.

We primarily follow the atomic data recommendations laid out in Table 4 of \cite{Berg15}, with two exceptions. First, we use the updated transition probabilities of \cite{Rynkun19} for S$^+$. Second, we use the collision strength data of \cite{Aggarwal99} for O$^{++}$. We recalculate the UV diagnostics using archival emission line strengths, however the collision data from \cite{Storey14}, recommended by \cite{Berg15}, do not cover the ultraviolet transitions (i.e. \oiiiuvfull). We therefore choose to use the collision data of \cite{Aggarwal99} for consistent comparisons between results. 

Ionic abundances relative to hydrogen are calculated using 
\begin{equation}
    \frac{N(X^i)}{N(H^+)} = \frac{I_{\lambda(i)}}{I_{\hbeta}} \frac{j_{\hbeta}}{j_{\lambda(i)}} .
\end{equation}
The emissivity coeffecients $j_{\lambda(i)}$ are calculated with the relevant temperature and density measurements described above, using the \texttt{PYNEB} function \texttt{getIonAbundance} \citep{pyneb_luridiana15}.
As with the temperature and density measurements, we account for uncertainties by randomly sampling emission line fluxes and electron temperatures from Gaussian distributions defined with mean equal to our measured flux/temperature, and standard deviation equal to the uncertainty on that parameter. 
We sample 300 iterations, and use the median and standard deviation of the resulting distribution to estimate the final abundance and uncertainty.

We use our measured $T_e(\textrm{O}^+)$ and $T_e(\textrm{O}^{++})$ values to calculate the oxygen abundance, and we use $T_e(\textrm{N}^+)$ calculated using Equation \ref{eq:n2o3} to estimate the nitrogen abundance for \banana. 
We also use the upper limit on the N$+$ temperature to calculate a lower limit on the nitrogen abundance. This limit is consistent with our abundance calculated using Equation \ref{eq:n2o3}.
We do not attempt to calculate a sulfur abundance, since we have neither a measurement of the electron temperature $T_e(\textrm{S}^{++})$, nor a measurement of the \siiidbl ~doublet. As mentioned earlier, an additional observation with the JWST NIRSpec G235H grating could measure the \siiidbl ~ doublet, enabling a measurement of the S$^{++}$ electron temperature and direct measurement of the sulfur abundance.

For \crab, we use our measured upper limit on $T_e(\textrm{O}^{++})$ and the calculated limit on $T_e(\textrm{O}^+)$ using the empirical temperature relation of \cite{G92} to calculate the total oxygen abundance. We do not attempt to extract limits on the abundance of other elements for this galaxy, as the requisite auroral lines are not detected.

We assume that the total oxygen abundance is given by $\textrm{O}/\textrm{H} = \textrm{O}^+/\textrm{H}^+ + \textrm{O}^{++}/\textrm{H}^+$. 
We do have a measurement of \oisixthree, which could provide a constraint on the abundance of oxygen in the lowest ionization state. However, the \oisixthree ~ line is partially cut off by the NIRSpec detector gap, and is not necessarily cospatial with the HII regions from which the other auroral lines and Balmer lines originate. We therefore choose not to include this line diagnostic because it only covers a small portion of the galaxy.
Additionally, the contribution from [OI] is expected to be low. \cite{Berg20} find that it contributes only 3\% of the total oxygen abundance in their sample of local HII regions, corresponding to a total change of $< 0.02$ dex, which is less than our measurement uncertainties. 

We make the assumption that N/O $\simeq$ N$^{+}$/O$^{+}$, based on the similar ionization and excitation energies of these ions \citep{Peimbert67}. Previous work has found this to be valid within $\sim 10\%$ \citep{Nava06,Amayo21}. This assumption offers sufficient precision for the present analysis.

We use the ICF presented in \cite{Izotov06} to calculate the total abundance of Ar/H for \banana. Following their recommendations, we linearly interpolate between the ``intermediate" and ``high" metallicity ICF presented in Equation 22 of \cite{Izotov06}, given our measured direct oxygen abundance falls in this range.
The ionization energy of Ar$^{++}$ overlaps both the intermediate and high ionization zones. We choose to use the high-ionization temperature $T_e(\textrm{O}^{++})$ when calculating the Ar$^{++}$ abundance, since we have a direct measurement of $T_e(\textrm{O}^{++})$.

\begin{table}[t]
    \centering
    \caption{Temperature and Abundance Measurements}
    \begin{tabular}{c c c}
    \hline \hline 
       Target  & \banana & \crab \\
       \hline 
        $T_e(\textrm{O}^{++})_{\textrm{meas.}}$ (K) & $12200 \pm 400$ & $ < 11700 $ \\ 
        $T_e(\textrm{O}^+)_{\textrm{meas.}}$ (K) & $13800 \pm 1000$ & $ < 5200$ \tablenotemark{*} \\ 
        $T_e(\textrm{N}^+)_{\textrm{meas.}}$ (K) & $< 12200$ & $< 22400$ \\ 
        $n_{e, \textrm{meas.}} (\textrm{cm}^{-3})$  & $110\pm 130$ & $600^{+400}_{-300}$ \\
        $n_{e, \textrm{used}} (\textrm{cm}^{-3})$  & $100$ & $600$ \\
        $T_e(\textrm{O}^+)_{\textrm{TT}}$ (K) & $11600 \pm 300$ & $ < 11500$ \\ 
        $T_e(\textrm{N}^+)_{\textrm{TT}}$ (K) & $11600 \pm 300$ & $< 11500$ \\ 
        $T_e(\textrm{S}^{++})_{\textrm{TT}}$ (K) & $11800 \pm 300$ & $< 11800$\\
        \hline 
        O$^+$/H$^+$ & $1.6 \pm 0.7 \times10^{-5}$  & $>2.2\times10^{-5}$ \\
        O$^{++}$/H$^+$ & $1.20 \pm 0.12 \times 10^{-4}$ & $> 8.6\times10^{-5}$ \\
        \hline
        N$^+$/H$^+$ & $> 1.5\times10^{-6}$ & -- \\
        N$^+$/H$^+$$_{\textrm{TT}}$ & $1.7\pm0.2\times10^{-6}$ & -- \\
        $\log$(N/O) & $ > -1.01$ & -- \\
        $\log$(N/O)$_{\textrm{TT}}$ & $ -0.95 \pm 0.2$ & -- \\
        \hline
        Ar$^{++}$/H$^+$ & $4.2 \pm 0.4 \times 10^{-7}$ & -- \\
        ICF(Ar) & $1.20\pm0.12$ & -- \\
        Ar/H & $5.0\pm0.7\times10^{-7}$ & -- \\
        log(Ar/O) & $-2.43\pm 0.08$ & -- \\
    \hline
        $12 + \log$(O/H) (Direct) & $8.13 \pm 0.03$ & $> 8.04$ \\
        $12 + \log$(O/H) (N2\halpha) & $8.44 \pm 0.06$ & $8.22 \pm 0.06$ \\
        $12 + \log$(O/H) (R23) & $8.42 \pm 0.09$ & $8.45 \pm 0.12$ \\
        \hline
    
    \end{tabular}

    \tablenotetext{}{\textbf{Note.} Electron temperatures and densities, and abundances calculated for both targets. Measured temperatures ($T_e(\textrm{X})_{\textrm{meas.}}$) are calculated from detected auroral lines. Upper limits are derived from undetected auroral lines. Temperatures denoted $_{\textrm{TT}}$ are calculated from $T_e(\textrm{O}^{++})_{\textrm{meas.}}$ using the empirical relations of \cite{G92}. Abundance measurements denoted $_{\textrm{TT}}$ are calculated using the relevant $T_e(X)_{\textrm{TT}}$ temperatures.}
    \tablenotetext{*}{This temperature limit relies on an uncertain rescaling of ground-based \oiibrightdbl ~flux measurements, and is likely too restrictive. We include it here for completeness only.}
    \label{tab:temp-abund}
\end{table}

\section{Results \& Discussion} \label{sec:results}

\begin{figure}[t]
    \centering
    \includegraphics[width=0.45\textwidth]{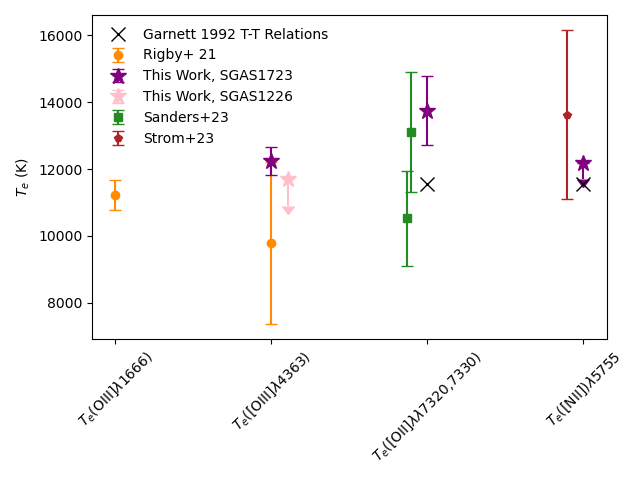}
    \caption{Electron temperatures from auroral line diagnostics for \banana~ are plotted as purple stars. \crab ~does not have detected auroral lines, an upper limit for the [OIII] temperature is plotted as a pink star. Comparison to \cite{Rigby21} measurements for \banana~ are shown as yellow circles, along with a calculation of the [OII], and [NII] temperature from the [OIII] temperature, using the temperature relations of \cite{G92}. While \banana~ has a detection of \siiiauroral, we do not have an available measurement of \siiidbl, which is needed to make a temperature measurement. Other high redshift O$+$ and N$+$ temperature measurements are shown from \cite{Sanders23_keck} at $z\sim2$ (green squares) and from the stacked sample of \cite{Strom23_Cecilia_lines} at $z = 2-3$ (red pentagon). Our measured temperatures are consistent with these other high redshift measurements.}
    \label{fig:temp}
\end{figure}

\begin{figure}[t]
    \centering
    \includegraphics[width=0.45\textwidth]{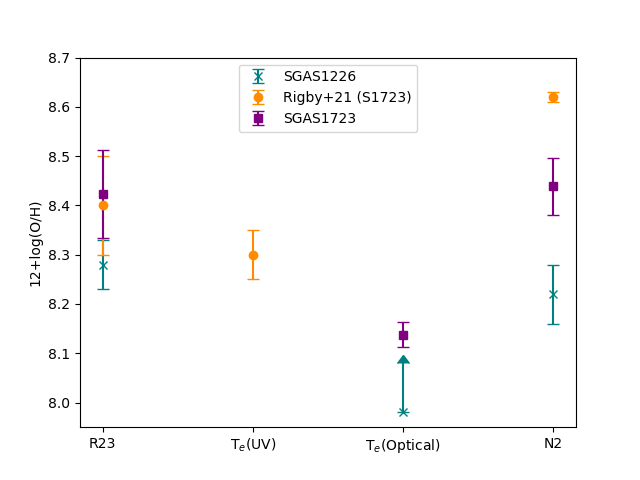}
    \caption{Oxygen abundance measurements $12+\log(\textrm{O}/\textrm{H})$ from both temperature and strong line diagnostics. The direct temperature-based measurement shown for \crab ~ is a lower limit, as no auroral lines are clearly detected in that target. For \banana, the direct abundance measurement is 0.3 dex lower than the strong-line diagnostics. The strong-line diagnostics R23 and N2 are self-consistent for both targets. Note that the direct abundance from \cite{Rigby21} utilizes the temperature measured from \oiiiuvfull, which is likely the cause of the offset with our measurement based on \oiiiauroral.}
    \label{fig:abund}
\end{figure}

\subsection{Multiple Auroral Line Detections}

We have measured fluxes for the auroral emission lines \oiiiauroral, \siiiauroral, and \oiiauroraldbl ~in the lensed galaxy \banana, and set an upper limit on the \niiauroral ~line flux.
This is the second detection of the auroral \siiiauroral ~line published for an individual galaxy outside the local universe after \cite{Rogers24}, suggesting that future JWST observations have a reasonable chance of detecting this feature in many more distant galaxies.
We have set upper limits for each of the auroral lines for \crab.
Using these emission line measurements, we have constrained the electron temperature and density within these two galaxies, and measured ionic abundances for \banana. 
Our measured values are reported in Table \ref{tab:temp-abund}.

Recently \cite{Strom23_Cecilia_lines} published deep, stacked spectra of $z\sim 2-3$ galaxies observed with JWST, including detections of three auroral lines: \niiauroral, \siiiauroral, and \oiiauroraldbl. 
The authors present fluxes of these auroral lines as a percentage of \halpha, offering a valuable comparison point for undetected lines in the present data. 
For \banana, we find an upper limit for the undetected \niiauroral~ line of $< 4.5\times 10^{-18} \textrm{ erg s}^{-1}\textrm{cm}^{-2}$ (see Table \ref{tab:lines}). 
\cite{Strom23_Cecilia_lines} present a 68\% confidence range of \niiauroral ~percentages of \halpha ~of [0.15 -- 0.25]\%. 
Applying this to our measured \halpha ~flux, we find an expected 68\% confidence range of $7.5 - 12.7 \times 10^{-18}\textrm{ erg s}^{-1}\textrm{cm}^{-2}$, which is greater than our measured upper limit. 
However, our measured \niib/\halpha ~ ratio for \banana ~ of $3.8\pm 0.4$\% is about half that reported in \cite{Strom23_Cecilia_lines}. If we assume that this is due to lower N$^+$/H in \banana ~(either from lower N/H or higher ionization parameter relative to the CECILIA stack), the \niiauroral ~line strength could fall by a similar margin at fixed electron temperature. This would bring the expected range to $3.8 - 6.4 \times 10^{-18}\textrm{ erg s}^{-1}\textrm{cm}^{-2}$, which is consistent with our measured upper limit on \niiauroral.

Our measured flux for \siiiauroral ~ is consistent with the 68\% confidence interval presented in \cite{Strom23_Cecilia_lines}.
For \siiiauroral, our IFU data only covers one clump of the full lensed arc. Thus for accurate comparison, we calculate the \halpha ~strength in that clump alone, finding a value of $212 \pm 2 \times 10^{-17} \textrm{ erg s}^{-1}\textrm{cm}^{-2}$, giving a ratio of \siiiauroral/\halpha ~of $0.4 \pm 0.1$\%. 
This is consistent with the range measured by \cite{Strom23_Cecilia_lines}.

Our measurement of \oiiauroraldbl ~is slightly below the range given in \cite{Strom23_Cecilia_lines}. We find [OII] ratios of $0.69\pm0.09$\% and $0.64\pm 0.08$\% for the blue and red lines, respectively. 
The other existing measurements of \oiiauroraldbl ~outside the local universe come from \cite{Sanders23_keck,Sanders23_ceers}. Both of these studies present line strengths which are $\sim 1.5-2$\% of \halpha ~ in a total of four galaxies, slightly above of the range presented in \cite{Strom23_Cecilia_lines}. 
Together, these measurements highlight the range of abundances present in galaxies at Cosmic Noon.

For \crab, we did not detect auroral lines. We set upper limits on the line strengths and use those upper limits to calculate upper limits on electron temperature, but do not attempt a direct metallicity measurement.
We compare our measured upper limits to the ranges seen in other cosmic noon galaxies from \cite{Strom23_Cecilia_lines}. 
For \oiiauroraldbl, we calculate expected $1\sigma$-low line fluxes of $4.4\times10^{-18}$ and $3.5\times 10^{-18} \textrm{ erg s}^{-1}\textrm{cm}^{-2}$, respectively. 
These expected values are brighter than our measured upper limits of $0.9\times10^{-18} \textrm{ erg s}^{-1}\textrm{cm}^{-2}$. 
Similarly, for \siiiauroral, we find a $1\sigma$-low expected value of $1.2\times10^{-18} \textrm{ erg s}^{-1}\textrm{cm}^{-2}$, slightly brighter than our upper limit of $1.1\times10^{-18} \textrm{ erg s}^{-1}\textrm{cm}^{-2}$. 
This indicates that we might have expected to detect these lines.
The fact that we do not detect these auroral lines indicates that this galaxy may be at a somewhat lower temperature (and thus higher metallicity) than expected based on strong line diagnostics.

\subsection{Electron Temperatures of Multiple Ionization Zones}

We use the auroral line flux measurements and upper limits to calculate electron temperatures.
We compare our results for \banana~ to previously published temperature measurements from \cite{Rigby21}, as plotted in Figure \ref{fig:temp}. 
We recomputed the literature temperatures based on the reported measured line fluxes with our reddening correction applied, and using the same atomic data as in our temperature calculations for consistency.
Our temperature measurement with the \oiiiauroral ~line diagnostic is consistent with the literature value within the $1\sigma$ uncertainties. 
We also find that our measured temperature is consistent with the \oiiiuvfull ~diagnostic previously published within $1\sigma$. 
The key improvement from the TEMPLATES data is the significant reduction in uncertainties. The \oiiiauroral ~temperature uncertainty has been reduced by a factor 6. This highlights the unique of JWST to deliver precise electron temperature diagnostics with modest observing time requirements.

\cite{Strom23_Cecilia_lines} only report the temperature for the low ionization gas, based on the \niiauroral ~line. 
Our measured upper limit on $T_e(\textrm{N}^+) < 12300$ is consistent with their stacked value ($ T_e = 13630 \pm 2540$ K).
Our measured \niiauroral ~flux limit is low compared to their sample, however we also find a lower \niia ~flux. When the difference in \niia ~is accounted for, this discrepancy disappears.

The only other high redshift temperature measurements for low ionization O$^+$ comes from \cite{Sanders23_keck,Sanders23_ceers}.
Their temperatures are consistent with our measured $T_e(\textrm{O}^+)$ within uncertainties.

\subsection{Oxygen Abundances From Direct and Strong Line Methods}

Using our measurements of the electron temperature from the \oiiiauroral ~and \oiiauroraldbl ~lines, we measure a direct metallicity for \banana~ of $12+\log(\textrm{O}/\textrm{H}) = 8.15 \pm 0.03$. 
This value is $\sim0.15$ dex below the direct oxygen abundance reported in \cite{Rigby21}, which is based on the \oiiiuvfull ~line. 
Dust will effect the UV lines more significantly than the optical, which could cause the discrepancy in these two abundance results.
Our estimate of the dust reddening correction in this galaxy is $E(B-V) = 0.05 \pm 0.01$, as described in Section \ref{sec:redcorr}. To test if dust could plausibly cause the observed offset, we artificially increased the reddening correction iteratively until the two measurements were consistent within uncertainties.
We find that a value of $E(B-V)\sim 0.076$ brings the UV and optical measurements into agreement within uncertainties, $2.6\sigma$ greater than our measured value.
The shape of the reddening curve in the UV carries additional uncertainty. Repeating the reddening correction test with the reddening curve of \cite{Fitzpatrick99} yields a value of $E(B-V) = 0.07$ to get consistent UV and optical abundance results. Meanwhile, the curve of \cite{Misselt99}, measured from the Large Magellanic Cloud, gives $E(B-V)=0.065$ with consistent UV and optical abundances, within $1.5\sigma$ of our measured reddening. We thus conclude that the observed offset is likely due to systematic uncertainty in the UV reddening correction.

We also computed the metallicity based on strong line diagnostics, namely R23 (R23 $ = \{\oiibrightdbl + \oiiidbl\} / \hbeta$) and N2 (N2 $ = \niib / \halpha$). 
These bright line indicators are plotted in Figure \ref{fig:abund} along with the direct measurements from this work and \cite{Rigby21}.
The R23 diagnostic gives an oxygen abundance $12+\log(\textrm{O}/\textrm{H}) = 8.42 \pm 0.09$ for \banana, using the relation presented in \cite{Kewley19_rev}.
The R23 abundance estimate is higher than the direct method abundance ($12+\log(\textrm{O}/\textrm{H}) = 8.14 \pm 0.03$) by $\sim0.3$ dex, however our R23 abundance is consistent with the R23 abundance calculated in \cite{Rigby21}. 
Discrepancies between the direct and strong-line abundances have been noted previously \citep[e.g.][]{Stasinska05,Kewley&Ellison08,LopezSanchez12}. 
The R23 indicator notably has two possible branches in many formulations, which could explain the discrepancy seen here. For example, the lower branch presented in \cite{Ly2016} gives a metallicity of $12+\log(\textrm{O/H}) \sim 8.15$, consistent with our direct measurement. Meanwhile the higher branch of that same paper gives a metallicity of $12+\log(\textrm{O/H}) \sim 8.45$, consistent with our strong line diagnostics. Another possible cause could be evolution with redshift, which \cite{Garg23} suggest could be driven by redshift evolution of the ionization parameter. While the \cite{Kewley19_rev} relation attempts to correct for ionization parameter, the effect of redshift evolution on metallicity indicators will require additional study.

Our N2 metallicity indicator gives an oxygen abundance of $12+\log(\textrm{O/H}) = 8.44 \pm 0.06$ using the relation of \cite{Kewley19_rev}, consistent with our measurement from R23.
The N2 metallicity calculated in \cite{Rigby21} is higher than our measurement by $\sim0.2$ dex. 
The N2 metallicity indicator in that paper was calculated from Gemini GNIRS data that only covered a small portion of the full arc. This minimal spatial coverage could be contributing to the offset between that work and our result.  
Interestingly, using the empirical calibration in Equation 1 of \cite{PettiniPagel04} gives an N2 metallicity measurement of $12+\log(\textrm{O}/\textrm{H}) = 8.09 \pm 0.03$, consistent with our direct measurement within uncertainties.

\cite{Curti17} note that at higher metallicities (starting around $12+\log(\textrm{O}/\textrm{H}) \approx 8.3$), contamination from the \feii ~line can artificially inflate the measured \oiiiauroral ~flux, decreasing the inferred direct metallicity. We test for any contribution from \feii ~ by including it as an extra component in fitting the \oiiiauroral ~line. We find that the \feii ~component fits to zero flux, indicating that the \oiiiauroral~line flux is not contaminated. 
Further, we see no evidence of other iron lines present in the spectrum, indicating that \feii ~contamination is unlikely for this source. 

We also calculate strong line oxygen abundance diagnostics for \crab, which are reported in Table \ref{tab:temp-abund}, and shown in Figure \ref{fig:abund}. 
These metallicity values are consistent with previous estimates \citep{Wuyts12,Saintonge13}.
Our metallicity from R23 is slightly higher than what we measure from N2, though the two are consistent within $2\sigma$. The discrepancy is likely due to systematic effects introduced from attempting to match the archival Keck NIRSPEC data to the JWST NIRSpec IFU data. The N2 diagnostic utilizes only the JWST data, making that the more reliable metallicity indicator in this case.
The lack of detected auroral lines prevents us from measuring the metallicity via the direct method for \crab.
However, we can set a lower limit of $12+\log(\textrm{O/H}) \geq 8.04$, less than and thus consistent with our strong line measurements. As mentioned previously, the \oiiiauroral ~line may be just below the detection threshold for the current observations. Deeper data covering this feature could plausibly yield a detection in the future.

\subsection{Nitrogen Enrichment}

For \banana, we calculate the direct nitrogen abundance. 
We do this in two ways, first using the calculated upper limit on the \niiauroral ~line, and second using the O$^{++}$ temperature calculated from the \oiiiauroral ~line using Equation \ref{eq:n2o3}. We report both N$^+$/H$^+$ abundance estimates in Table \ref{tab:temp-abund}. 
We find that the upper limit set by our non-detection of \niiauroral~ is slightly lower ($1.2\sigma$) than the measured O$^+$ temperature.
While there is considerable scatter in the $T_e(\textrm{N}^+$)-$T_e(\textrm{O}^+)$ relation in local galaxies, \cite{Berg20} find that the oxygen temperatures are typically slightly higher ($\sim 1000-2000$~K) than the nitrogen temperatures. Our measurements are consistent with this picture. 
We use the O$^{++}$ temperature and Equation \ref{eq:n2o3} to estimate the low-ionization electron temperature for our calculations of the nitrogen abundance. 
While the O$^+$ temperature offers a direct probe of the low-ionization electron temperature, this diagnostic is subject to several major systematic uncertainties in this study. The ratio of \oiibrightdbl/\oiiauroraldbl ~is strongly dependent on the reddening correction. In this case, the \oiibrightdbl~and \oiiauroraldbl ~lines are measured by different instruments, adding an additional source of systematic uncertainty.
Deeper observations would be needed to detect the \niiauroral~ line and refine these nitrogen abundance measurements. 

\begin{figure}
    \centering
    \includegraphics[width=0.49\textwidth]{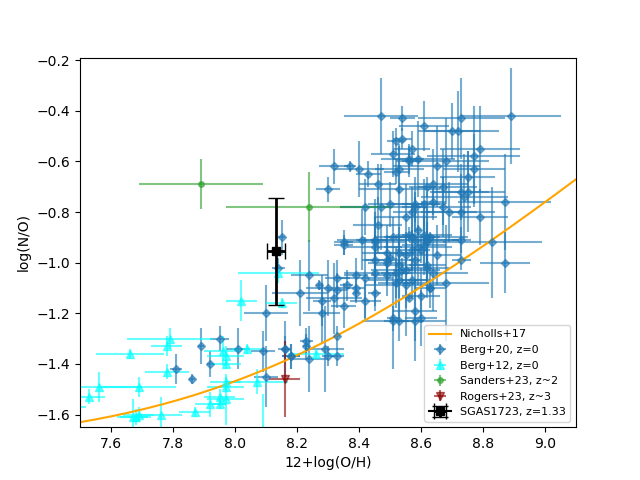}
    \caption{N/O vs O/H from temperature-based abundance measures is plotted for \banana~ (black square), alongside local HII regions (\cite{Berg20}, blue diamonds), local dwarf galaxies (\cite{Berg12}, cyan triangles) and the nebular scaling relation calculated in \cite{Nicholls17} (orange line). Other high-redshift points from \cite{Sanders23_keck} are demarcated with green circles, and from \cite{Rogers24} with a red triangle. The nitrogen abundance for \banana~ is calculated using the O$^{++}$ electron temperature (see Section \ref{sec:abundcalc}. Our measurement of the N/O ratio is consistent with local observations of star forming galaxies, though we cannot rule out enhanced nitrogen abundance with our current uncertainties.}
    \label{fig:oh-no}
\end{figure}

As a consistency check, we calculate the N/O abundance using our measured O$^+$ temperature. We find that the N/O($T_e$(O$^+)) = -1.1 \pm 0.2$, consistent with our quoted value using Equation \ref{eq:n2o3} within $1\sigma$ uncertainties. 

We calculate the total N/O ratio as discussed in Section \ref{sec:abundcalc}.
We compare the N/O ratio for \banana~ to local HII regions from \cite{Berg20}, as well as nearby ($z\sim0$) galaxies from \cite{Berg12} in Figure \ref{fig:oh-no}, and find that this target is consistent with local measurements. 
Our measurement is also slightly higher than the scaling relation for gaseous nebulae calculated by \cite{Nicholls17}, though still within $2\sigma$.

Interestingly, some previous measurements of N/O in $z\sim 2$ galaxies have shown significantly elevated nitrogen abundances compared to local galaxies \citep{Sanders23_keck}, while another measurement of a galaxy at $z\sim 3$ has shown a nitrogen abundance consistent with $z\sim 0$ galaxies \citep{Rogers24}.
While a number of factors could influence the elevated nitrogen abundances of the galaxies studied in \cite{Sanders23_keck}, factors such as a top-heavy stellar initial mass function, leading to a larger number of Wolf-Rayet stars shedding nitrogen into the interstellar medium, have been proposed to generate this large amount of nitrogen quickly \citep[e.g.,][]{Henry2000,Brinchmann2008,Kobayashi24}. 
An alternative explanation suggests that inflowing low-metallicity gas can dilute the oxygen abundance, driving increased N/O without requiring the presence of Wolf-Rayet stars \citep[e.g.,][]{Koppen2005,Amorin2010,Andrews2013}.

We see no compelling evidence for this increased nitrogen fraction in \banana, though we note that the best-fit value of $\log(\textrm{N/O})$ does fall at the high end of the range seen in $z\sim 0$ galaxies reported in \cite{Berg12}.
A detection of the \niiauroral ~ line would improve the measurement of N/O for this galaxy and clarify whether the nitrogen abundance is discrepant compared to $z\sim 0$ low-metallicity galaxies.
Ultimately a larger sample of direct chemical abundance measurements will be needed in order to ascertain whether the overabundance of nitrogen in distant galaxies is typical or unusual. 
Additionally, we note that a variety of systematic effects could be influencing the observed discrepancies in these $z\sim 1-3$ galaxies. For example, \cite{Sanders23_keck} use the \oiiauroraldbl ~lines as their primary temperature diagnostic for the low ionization gas, while this study and \cite{Rogers24} use the \oiiiauroral ~line and the relation of \cite{G92} to calculate the low ionization gas temperature. Larger self-similar samples will help to determine the history of nitrogen synthesis in galaxies.

\subsection{Ar/O Abundance}

We detect the \ariii ~line for \banana, allowing us to make a direct measurement of the Ar/H abundance, as well as the ratio of Ar/O. As reported in Table \ref{tab:temp-abund}, we find an Ar/O ratio of $\log(\textrm{Ar}/\textrm{O}) = -2.44 \pm 0.07$, consistent with the solar abundance $\log(\textrm{Ar}/\textrm{O}) = -2.37 \pm 0.11$ \citep{Asplund21}.

Recently, \cite{Rogers24} reported a direct Ar/O abundance of $\log(\textrm{Ar}/\textrm{O}) = -2.71 \pm 0.09$ at $z\sim 3$, noting that this is significantly lower than the solar abundance. They use this measurement to argue that this galaxy has been enriched mainly by core-collapse supernovae, as fewer Type Ia supernovae would have time to develop by $z\sim 3$. 
Our measurement for \banana ~being consistent with solar Ar/O abundance suggests that it is further along in its evolution, such that enough time has elapsed for enrichment by Type Ia supernovae.
Additional direct argon abundance measurements at high-$z$ will provide further nuance to the buildup of $\alpha$-elements over cosmic time.

\section{Conclusions} \label{sec:conclusion}

We report the detection of multiple auroral lines in the gravitationally lensed galaxy \banana ~($z=1.3292$). 
Using these lines, we calculate the temperature of both the high and low ionization regions of oxygen gas, finding $T_e \sim 12000$ K in both regions.
These temperatures enable calculation of the total oxygen abundance via the direct method, yielding a value of $12+\log(\textrm{O/H}) = 8.15\pm0.03$. 
We also use the low ionization temperature to calculate the nitrogen abundance in this galaxy, finding it is consistent with local HII regions.
We do not detect auroral line emission in the lensed galaxy \crab. 

This study focused on the coadded spectra, collapsing all of the information from these lensed galaxies into a single spectrum to maximize signal to noise. 
The TEMPLATES NIRSpec IFU observations present an opportunity to spatially resolve these emission lines. 
Analysis of the spatially resolved abundances in \banana~ will be presented in a forthcoming companion paper (Olivier et al., in prep).

\begin{acknowledgments}
We thank Allison Strom and Gwen Rudie of the CECILIA team for collegial discussions that strengthened this paper. 
This work was based on observations taken with the NASA/ESA/CSA JWST.
JWST is operated by the Space Telescope Science Institute under the management of the Association of Universities for Research in Astronomy, Inc., under NASA contract no. NAS 5-03127.
These observations were taken as part of JWST DD-ERS program 1355.
Support for JWST program 1355 was provided by NASA through a grant from the Space Telescope Science Institute, which is operated by the Association of Universities for Research in Astronomy, Inc., under NASA contract NAS 5-03127.
The JWST data analyzed in this paper were obtained from the Mikulski Archive for Space Telescopes (MAST) at the Space Telescope Science Institute. The specific observations analyzed can be accessed via \dataset[DOI]{http://dx.doi.org/10.17909/e9p4-xm44}
BW acknowledges support from NASA under award number 80GSFC21M0002.
\end{acknowledgments}

%% To help institutions obtain information on the effectiveness of their 
%% telescopes the AAS Journals has created a group of keywords for telescope 
%% facilities.
%
%% Following the acknowledgments section, use the following syntax and the
%% \facility{} or \facilities{} macros to list the keywords of facilities used 
%% in the research for the paper.  Each keyword is check against the master 
%% list during copy editing.  Individual instruments can be provided in 
%% parentheses, after the keyword, but they are not verified.

\vspace{5mm}
\facilities{JWST}

%% Similar to \facility{}, there is the optional \software command to allow 
%% authors a place to specify which programs were used during the creation of 
%% the manuscript. Authors should list each code and include either a
%% citation or url to the code inside ()s when available.

\software{astropy \citep{Astropy13,Astropy18,Astropy22},  
          scipy \citep{2020SciPy-NMeth},
          matplotlib \citep{Matplotlib_Hunter07}
          }

%% Appendix material should be preceded with a single \appendix command.
%% There should be a \section command for each appendix. Mark appendix
%% subsections with the same markup you use in the main body of the paper.

%% Each Appendix (indicated with \section) will be lettered A, B, C, etc.
%% The equation counter will reset when it encounters the \appendix
%% command and will number appendix equations (A1), (A2), etc. The
%% Figure and Table counter will not reset.

\appendix

\section{Measuring the [\textrm{S}\textsc{III}] Auroral Line Flux with Local Continuum Subtraction} \label{ap:s3local}

As mentioned in Section \ref{sec:linemeasure}, the \siiiauroral~line produces only an upper limit on the flux when using our standard continuum subtraction and line fitting procedure in \banana. 
However, visual inspection of the IFU data cube reveals that an emission line is present at the expected wavelength. 
This line is only partially visible due to the NIRSpec detector gap.
The lack of an apparent emission line in the final coadded spectrum is due to increased noise from detector edge pixels within our spectral extraction aperture. 

\begin{figure}
    \centering
    \includegraphics[width=0.5\linewidth]{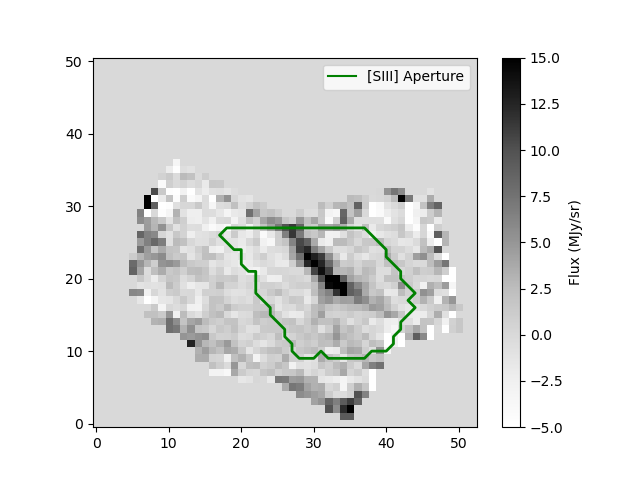} 
    \includegraphics[width=0.45\linewidth]{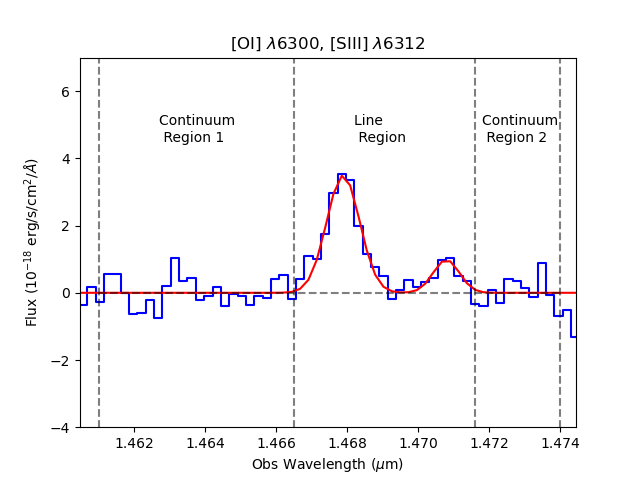} 
    \caption{The auroral \siiiauroral ~line appears near the NIRSpec detector gap in observations of \banana, causing only part of the arc to be visible. The left panel shows a sum of 10 IFU cube slices centered on \siiiauroral, in the JWST pipeline units of MJy/sr. Because the edge spaxels tend to be noisier than others, we use a smaller extraction aperture (green line) to measure the flux of this line that avoids unnecessary edge spaxels. The resulting continuum-subtracted spectrum is shown in the right panel (blue line). We use a local continuum fitting technique for this line, utilizing two continuum regions adjacent to the emission lines of interest. Our fit to the \oisixthree ~and \siiiauroral ~lines is shown in red. This method gives us the most reliable flux measurement for this faint auroral line on the edge of the detectors.}
    \label{fig:s3local}
\end{figure}

In an effort to recover a line flux for \siiiauroral, we create a partial extraction aperture that avoids the noisy detector edge pixels, as outlined in red in Figure \ref{fig:s3local}.
This follows our regular aperture along the southern portion of the arc for consistency with our other measurements.
We make no attempt to correct the flux for the fact that only part of the galaxy is observed, since we do not have the \siiidbl~lines required to make a temperature measurement for [SIII]. 
If future observations measure the \siiidbl~lines, a more detailed treatment of this line flux will be warranted.

To measure the \siiiauroral ~line flux, we employ a local continuum subtraction technique using \texttt{specutils}. 
We select continuum regions bracketing the \oisixthree ~and \siiiauroral ~lines, as shown in Figure \ref{fig:s3local}. 
This process preserves the flux of the \siiiauroral ~ line, and allows us to fit it with a Gaussian rather than merely setting an upper limit. 
We report this flux in Table \ref{tab:lines}.

\bibliography{bib}{}
\bibliographystyle{aasjournal}

%% This command is needed to show the entire author+affiliation list when
%% the collaboration and author truncation commands are used.  It has to
%% go at the end of the manuscript.
%\allauthors

%% Include this line if you are using the \added, \replaced, \deleted
%% commands to see a summary list of all changes at the end of the article.
%\listofchanges

\end{document}